\documentclass[pra,aps,twocolumn,epsfig,nofootinbib,superscriptaddress]{revtex4}

\usepackage{amsmath}
\usepackage{amssymb}
\usepackage[dvips]{graphicx}
\usepackage{dcolumn}
\usepackage{bm}
\usepackage[driverfallback=dvips]{hyperref} 
\usepackage{setspace}
\usepackage{amsfonts}
\usepackage{rotating}
\usepackage{makeidx}
\usepackage{amsthm}
\usepackage{bbold}
\usepackage{float}

\begin{document}

 \title{Anyonic quantum multipartite maskers in the Kitaev model \footnote{Phys. Rev. A 109, 032421 (2024)}}

\author{Yao Shen}
\affiliation{School of Criminal Investigation, People's Public Security University
of China, Beijing 100038, China}

\author{Wei-Min Shang}
\affiliation{School of Science, Tianjin Chengjian University, Tianjin 300384, China}

\author{Chi-Chun Zhou}
\affiliation{School of Engineering, Dali University, Dali, Yunnan 671003, China}

\author{Fu-Lin Zhang}
\email[Corresponding author: ]{flzhang@tju.edu.cn}
\affiliation{Department of Physics, School of Science, Tianjin University, Tianjin 300072, China}

\date{\today}

\begin{abstract}
The structure of quantum mechanics forbids a bipartite scenario for masking
quantum information, however, it allows multipartite maskers. The Latin
squares are found to be closely related to a series of tripartite maskers.
This adds another item, significantly different from the original no-cloning
theorem, to the no-go theorems. On the other hand, anyonic excitations in
two dimensions exhibit exotic collective behaviors of quantum physics, and
open the avenue of fault-tolerant topological quantum computing.
Here, we give the Latin-square construction of Abelian and Ising anyons 
in the Kitaev model and %
study the maskable space configuration in anyonic space.
The circling and braiding of Kitaev anyons are masking operations on
extended hyperdisks in anyonic space. %
We also realize quantum information masking in a teleportation way in the
Kitaev Ising anyon model. %
\end{abstract}

	
 \maketitle
	
\section{Introduction}

Several no-go theorems have been founded in the study of quantum
information, such as no-cloning \cite{00,01}, no-deleting \cite{02},
no-broadcasting \cite{03,04}, and no-hiding theorems \cite{05}, which
describe the difference between the quantum and classical worlds.
Recently, Modi \emph{et al.} \cite{1} introduced another item, the
no-masking theorem, to the family of no-go theorems. That is, it is
impossible to mask an arbitrary state into bipartite quantum systems. %
However, the task can be achieved in multipartite scenarios, which reveals a
significant difference between the no-masking theorem and the original
no-cloning theorem.
Li \emph{et al.} \cite{2} presented a unified construction of tripartite
scenarios based on the Latin squares. %
Subsequently, probabilistic and approximate masking protocols are also
studied \cite{p}.

Quantum information masking is crucial to many quantum communication topics,
such as quantum secret sharing \cite{d1,d2} and quantum bit commitments \cite%
{c1,c2}. Optical experimental demonstrations of masking schemes have been
reported very recently \cite{optic1,optic2}. 
However, quantum masking in condensed-matter (many-body) systems is still
absent. %

This paper is devoted to the investigation of quantum information masking in
anyons, which are quasiparticles living in two-dimensional condensed-matter
systems. Anyons do not fit into the usual statistics of fermions and bosons,
but obey a new form of fractional statistics, closely related
with their famous braiding and fusion rules \cite{b1,b2,b3,b4,b5,b6,b7,L,MR,MR1}. %
The nontrivial topological properties of anyons in the Kitaev spin-lattice
model, opened the avenue of topological quantum computing \cite{K,K1,K2,K3,K4,Yu}.

The quantum information masking experiment is a technology that uses quantum
mechanics to mask and protect information. Although the quantum information
masking experiment has many advantages in theory, it still faces many
difficulties and challenges in practical applications. The Kitaev anyon
system provides different ideas and methods for solving these problems in
quantum information masking, making it a strong contender for experimental
implementation.

(1) The fragility of quantum states: Quantum states are very fragile and are
easily affected by noise and interference in the environment, leading to
information loss and errors. Therefore, how to effectively protect and
manipulate quantum states in quantum information masking experiments is a
huge challenge. The Kitaev anyon system is a topologically protected system
that can resist various forms of noise such as phase noise. There is an
energy gap between the ground-state energy and excited-state energy of anyon
systems, which can make the system immune to local errors. This makes anyon
systems robust for quantum computation and quantum information. \cite%
{K,K1,K2,K3}.

(2) Operation and control of qubits: In practical applications, the
operation and control of qubits require very precise and meticulous
attention, otherwise it will lead to information loss and errors. Therefore,
how to achieve precise and reliable control of quantum bits is also an
important challenge faced by quantum information masking experiments. The
toric code (or Kitaev surface code) provides a feasible and relatively
simple manipulation method \cite{K1,K4}.

(3) Generation and control of quantum entanglement: Quantum entanglement is
one of the key elements for realizing quantum information masking. However,
in practical experiments, generating and controlling high-quality quantum
entanglement remains a technical challenge. The nonlocal topological
properties of particles in the Kitaev model provide a potential solution to
this problem. The topological properties describe the overall properties of
the system rather than the properties of individual particles. In addition,
the ground state of a system of host non-Abelian anyons is usually highly
entangled \cite{K,K1,K2,K3}.

(4) The masking scheme should be universal for quantum circuits, and a
typical example, Ising anyons, are not intuitively universal. In the schemes
of Li \emph{et al.} \cite{2}, Latin squares are used to realize masking in
normal Hilbert spaces. 
But the anyonic space is composed of the direct product of each Hilbert
subsystem spaces and the fusion Hilbert space of all charges,
\begin{equation}
H_{AB}^{c}=\underset{ab}\oplus H_{A}^{a}\otimes H_{B}^{b}\otimes V_{ab}^{c},
\end{equation}
where $H_{A}^{a}$ is the space of charge $a$ in the Hilbert subsystem space $%
A$, and the space $V_{ab}^{c}$ is the fusion space containing the fusion
rules (details are shown in the following). In the process of fusion (in the
space $V_{ab}^{c}$), the braiding of charges induces the topological
properties and causes entanglement. The braiding process of Ising anyons not
only generates additional phases, but also accompanies the exchange of
particles (the creation and the annihilation of new particles---see Figs. %
\ref{fig3} and \ref{fig4}). It is intriguing whether they will interfere
with the masking process. Our work demonstrates that masking can still be
achieved, even in the case of Ising anyons.

In the present work, we construct the Latin-square masking scenarios of
Abelian anyons and Ising anyons (the simplest non-Abelian anyons) in the
Kitaev model. %
The braiding operations on anyons manipulate the states in the maximal
maskable space, which are extended hyperdisks.
This maskable space configuration of anyons allows a series of masking
schemes in many-body systems. We also demonstrate the process of masking
based on teleportation in the Kitaev Ising anyon model. 

This paper is organized as following: 
In the next section, we present the masking protocol in a system of Abelian
anyons using the Latin-square construction.
Section \ref{Ising} shows the process of masking in non-Abelian Ising
anyons, which has an explanation based on quantum teleportation.
Finally, the last section presents a summary.

\section{The Kitaev model and the Abelian $1/2$-anyon}

The Kitaev model is a honeycomb spin-lattice model in which each $1/2$ spin
is located on the vertex of the hexagon. Fermions and $Z_{2}$ vortices are
the excitation states of this exactly solvable model. The excitations are
divided into two cases, Abelian anyons and non-Abelian anyons, and
further-more $16$ types of statistics with Chern number $c$ mod $16$. Bosons
and gapped fermions are Abelian anyons which have an even Chern number $c$.
Vortices and gapless fermions are non-Abelian anyons whose Chern numbers are
odd \cite{K,K1}.

\subsection{Review of the Abelian $1/2$-anyon and braiding rules}

\begin{figure}[tbp]
\includegraphics[scale=0.8]{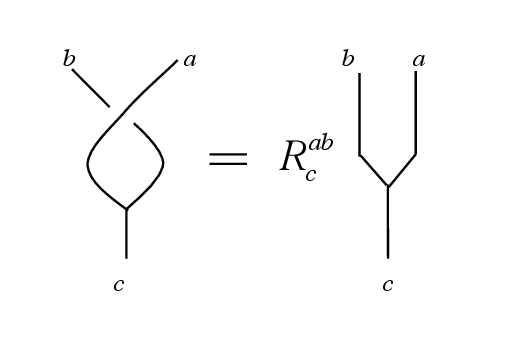}
\caption{The braiding operation of $a$ and $b$. The fusion of $a$ and $b$ is
$c$.}
\label{fig1}
\end{figure}

Wilczek \cite{a1,a2} first introduced the Abelian anyons which are
represented by the braiding group. He pointed out that the braiding of
different quasiparticles for one circle caused a Aharonov-Bohm phase $\exp
(i2\pi k\alpha )$ ($\alpha $ is the statistical parameter and $k$ is the
winding number) \cite{a1,a2,a3}. The Abelian anyons have four superselection
sectors: $\mathit{1}$ (the vacuum), vortices $e$ and $m$, and fermion $%
\varepsilon $ (we ignore another case in which the vortices are two mutual
antiparticles). There are two cases of Abelian anyons: One is the Chern
number $c=0,8$, and the other is $c=\pm 4$. Both of them are mod $16$ with
different topological spins and Frobenius-Schur indicators. Here, we
consider the simplest Abelian anyons (called Abelian anyons for short) with
a Chern number $c=0$, topological spin $\theta =1$, and Frobenius-Schur
indicator $\varkappa =1$. The fusion rules of the Abelian anyons give

\begin{equation}
e\times e=m\times m=1,\quad \varepsilon \times \varepsilon =1,
\end{equation}%
\begin{equation}
\varepsilon \times e=m,\quad \varepsilon \times m=e,\quad e\times
m=\varepsilon .
\end{equation}%
The Abelian anyons are mod $2$, which means that two of the same
quasiparticles annihilate to the vacuum. The braiding operation can be
represented by $R_{z}^{xy}$ as in Fig. \ref{fig1}.

\begin{figure}[tbp]
\includegraphics[scale=0.8]{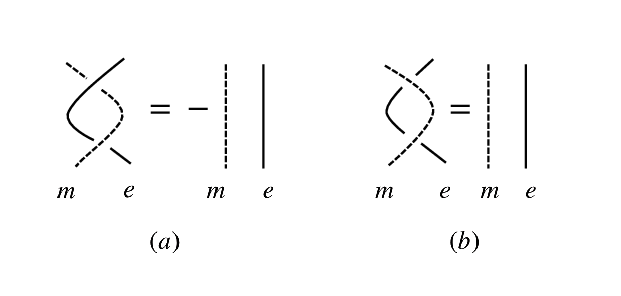}
\caption{The braiding (circling) of $e$ and $m$. (a) An $e$ particle
circling an $m$ particle counterclockwise corresponds to $R_{\protect%
\varepsilon }^{me}\cdot R_{\protect\varepsilon }^{em}=-1$. (b) There is no
braiding in this case, $R_{\protect\varepsilon }^{me}\cdot (R_{\protect%
\varepsilon }^{me})^{-1}=1$, where the two curves are separated. }
\label{fig2}
\end{figure}
In this case, as shown in Fig. \ref{fig2}, all associativity relations are
trivial and the braiding rules are
\begin{equation}
\begin{array}{c}
R_{\varepsilon }^{em}=1,\,R_{\varepsilon }^{me}=-1; \\
R_{m}^{e\varepsilon }=1,\,R_{m}^{\varepsilon e}=-1; \\
R_{e}^{\varepsilon m}=1,\,R_{e}^{m\varepsilon }=-1; \\
R_{1}^{ee}=R_{1}^{mm}=1,\,R_{1}^{\varepsilon \varepsilon }=-1;%
\end{array}
\label{eq:11}
\end{equation}%
The vortices $e$ and $m$ are bosons. The exchange of $e$ and $m$ is
different from that of $m$ and $e$ ($R_{\varepsilon
}^{em}=1,\,R_{\varepsilon }^{me}=-1$). With the help of the representation
of Gentile statistics, we gave a physical image for the exchange of
different Abelian anyons \cite{me}. Braiding different Abelian anyons
depends on the topology of the particles. Braiding two different
quasiparticles for one circle gives an additional phase $-1$ (e.g.,$%
R_{\varepsilon }^{em}R_{\varepsilon }^{me}=-1$ ) to the state, so these
kinds of anyons are called Abelian $1/2$-anyons ($\exp [i2\pi \cdot
(1/2)\cdot 1]=-1$).

\subsection{Masking protocol of the Abelian $1/2$-anyon}

According to the orthogonality of the matrix elements, Latin squares are
useful tools in quantum information masking. The matrix elements in each row
and column of the Latin squares are all different, which leads to the
disappearance of the cross terms during a trace. Two mutually orthogonal
Latin squares are those matrices whose products of the matrix elements at
the same locations are all different. Based on Theorem $2$ in Ref. \cite{2},
for a $d$-dimensional space, the quantum states can be masked in $\mathbb{C}%
^{d}\otimes \mathbb{C}^{d}\otimes \mathbb{C}^{d}$ systems. In the masking
protocol of the Abelian $1/2$-anyons, the space is four dimensional (four
superselection sectors $\left\{ \left\vert 1\right\rangle ,\left\vert
e\right\rangle ,\left\vert m\right\rangle ,\left\vert \varepsilon
\right\rangle \right\} $). We define three matrices $A$, $B$, $C$ in $%
\mathbb{C}^{4}\otimes \mathbb{C}^{4}\otimes \mathbb{C}^{4}$,
\begin{equation}
A\equiv \left(
\begin{array}{c}
1 \\
1 \\
1 \\
1%
\end{array}%
\right) \otimes \left(
\begin{array}{cccc}
1 & e & m & \varepsilon%
\end{array}%
\right) \quad \!\!\!\mathrm{or}\!\!\!\quad \left(
\begin{array}{c}
1 \\
e \\
m \\
\varepsilon%
\end{array}%
\right) \otimes \left(
\begin{array}{cccc}
1 & 1 & 1 & 1%
\end{array}%
\right) ,  \label{eq:a}
\end{equation}%
where $B$ and $C$ are two mutually orthogonal Latin squares in $\mathbb{C}%
^{4}$. The more ideal case is where $A$, $B$, $C$ are three mutually
orthogonal Latin squares in $\mathbb{C}^{4}$. The matrix $A$ can extend the
restrictive condition to Eq. (\ref{eq:a}). The masking protocol process maps
$\left\{ \left\vert j\right\rangle \left\vert \left\vert 1\right\rangle
,\left\vert e\right\rangle ,\left\vert m\right\rangle ,\left\vert
\varepsilon \right\rangle \right. \right\} $ to
\begin{equation}
\left\vert \psi _{j}\right\rangle =\frac{1}{2}\sum_{k}\left\vert
A_{jk}B_{jk}C_{jk}\right\rangle .  \label{eq:b}
\end{equation}%
So we have the encoding process as
\begin{equation}
\left\vert \Psi \right\rangle =\sum_{j}\alpha _{j}\left\vert \psi
_{j}\right\rangle =\frac{1}{2}\sum_{jk}\alpha _{j}\left\vert
A_{jk}B_{jk}C_{jk}\right\rangle .  \label{eq:c}
\end{equation}%
In this case,
\begin{equation*}
\mbox{Tr}_{AB}\left( \left\vert \Psi \right\rangle \left\langle \Psi
\right\vert \right) =\mbox{Tr}_{AC}\left( \left\vert \Psi \right\rangle
\left\langle \Psi \right\vert \right) =\mbox{Tr}_{BC}\left( \left\vert \Psi
\right\rangle \left\langle \Psi \right\vert \right) =\frac{I}{4},
\end{equation*}%
where $I$ is the identity matrix. The quantum information is stored in the
correlation of the tripartite system.

What calls for special attention is that the braiding of anyons (exchange)
and the braiding for circles (circling), which are equal to operations on
extended hyperdisks (which differ from hyperdisks on normal Hilbert space),
do not affect the masking process, because the state vectors and their
conjugations cancel out the phase factors attributed to the braidings. The
cases in point are the exchange of $B$ and $C$ or braiding $C$ around $B$
for one circle. In other words, quantum information masking is invariant
under the braiding operations of Abelian anyons. The details are given below.

$\mathit{Proof.}$We give one example and define
\begin{equation}
\begin{array}{c}
A=\left(
\begin{array}{cccc}
1 & e & m & \varepsilon \\
1 & e & m & \varepsilon \\
1 & e & m & \varepsilon \\
1 & e & m & \varepsilon%
\end{array}%
\right) ,:B=\left(
\begin{array}{cccc}
1 & e & m & \varepsilon \\
e & 1 & \varepsilon & m \\
m & \varepsilon & 1 & e \\
\varepsilon & m & e & 1%
\end{array}%
\right) , \\
C=\left(
\begin{array}{cccc}
1 & e & m & \varepsilon \\
\varepsilon & m & e & 1 \\
e & 1 & \varepsilon & m \\
m & \varepsilon & 1 & e%
\end{array}%
\right) .%
\end{array}%
\end{equation}

$B$ and $C$ are mutually orthogonal Latin squares. An arbitrary state $%
\alpha \left\vert 1\right\rangle +\beta \left\vert e\right\rangle +\gamma
\left\vert m\right\rangle +\delta \left\vert \varepsilon \right\rangle $ is
mapped to a tripartite system $\left\vert \Psi \right\rangle $,%
\begin{equation}
\begin{array}{c}
\rightarrow \frac{1}{2}[\alpha (\left\vert 111\right\rangle +\left\vert
eee\right\rangle +\left\vert mmm\right\rangle +\left\vert \varepsilon
\varepsilon \varepsilon \right\rangle ) \\
+\beta (\left\vert 1e\varepsilon \right\rangle +\left\vert e1m\right\rangle
+\left\vert m\varepsilon e\right\rangle +\left\vert \varepsilon
m1\right\rangle ) \\
+\gamma (\left\vert 1me\right\rangle +\left\vert e\varepsilon 1\right\rangle
+\left\vert m1\varepsilon \right\rangle +\left\vert \varepsilon
em\right\rangle ) \\
+\delta (\left\vert 1\varepsilon m\right\rangle +\left\vert em\varepsilon
\right\rangle +\left\vert me1\right\rangle +\left\vert \varepsilon
1e\right\rangle )].%
\end{array}%
\end{equation}%
\begin{equation}
\mbox{Tr}_{AB}(\left\vert \Psi \right\rangle \left\langle \Psi \right\vert )=%
\frac{1}{2}\sum_{jk}\left\vert \alpha _{j}\right\vert ^{2}\left\vert
C_{jk}\right\rangle \left\langle C_{jk}\right\vert =\frac{I}{4}.
\end{equation}%
With a similar operation, the other two partial traces are $\mbox{Tr}%
_{AC}\left( \left\vert \Psi \right\rangle \left\langle \Psi \right\vert
\right) =\mbox{Tr}_{BC}\left( \left\vert \Psi \right\rangle \left\langle
\Psi \right\vert \right) =I/4$. The mapping of Eq. (\ref{eq:c}) is indeed a
masking protocol.

Now we show the braiding (exchange) of $B$ and $C$. According to Eq. (\ref%
{eq:11}), braiding $B$ and $C$ gives
\begin{equation}
\begin{array}{c}
\rightarrow \frac{1}{2}[\alpha (\left\vert 111\right\rangle +\left\vert
eee\right\rangle +\left\vert mmm\right\rangle +\left\vert \varepsilon
\varepsilon \varepsilon \right\rangle ) \\
+\beta (R^{e\varepsilon }\left\vert 1e\varepsilon \right\rangle +\left\vert
e1m\right\rangle +R^{\varepsilon e}\left\vert m\varepsilon e\right\rangle
+\left\vert \varepsilon m1\right\rangle ) \\
+\gamma (R^{me}\left\vert 1me\right\rangle +\left\vert e\varepsilon
1\right\rangle +\left\vert m1\varepsilon \right\rangle +R^{em}\left\vert
\varepsilon em\right\rangle ) \\
+\delta (R^{\varepsilon m}\left\vert 1\varepsilon m\right\rangle
+R^{m\varepsilon }\left\vert em\varepsilon \right\rangle +\left\vert
me1\right\rangle +\left\vert \varepsilon 1e\right\rangle )].%
\end{array}%
\end{equation}%
It can be proved that we still have
\begin{equation*}
\mbox{Tr}_{AB}\left( \left\vert \Psi \right\rangle \left\langle \Psi
\right\vert \right) =\mbox{Tr}_{AC}\left( \left\vert \Psi \right\rangle
\left\langle \Psi \right\vert \right) =\mbox{Tr}_{BC}\left( \left\vert \Psi
\right\rangle \left\langle \Psi \right\vert \right) =\frac{I}{4}.
\end{equation*}%
Exchanging $A$ and $C$, and $A$ and $B$ is similar.

Next, when we braid $C$ around $B$ for one circle, so we have
\begin{equation}
\begin{array}{c}
\rightarrow \frac{1}{2}[(\alpha \left\vert 111\right\rangle +\left\vert
eee\right\rangle +\left\vert mmm\right\rangle +\left\vert \varepsilon
\varepsilon \varepsilon \right\rangle ) \\
+\beta (R^{e\varepsilon }R^{\varepsilon e}\left\vert 1e\varepsilon
\right\rangle +\left\vert e1m\right\rangle +R^{\varepsilon e}R^{e\varepsilon
}\left\vert m\varepsilon e\right\rangle +\left\vert \varepsilon
m1\right\rangle ) \\
+\gamma (R^{me}R^{em}\left\vert 1me\right\rangle +\left\vert e\varepsilon
1\right\rangle +\left\vert m1\varepsilon \right\rangle
+R^{em}R^{me}\left\vert \varepsilon em\right\rangle ) \\
+\delta (R^{\varepsilon m}R^{m\varepsilon }\left\vert 1\varepsilon
m\right\rangle +R^{m\varepsilon }R^{\varepsilon m}\left\vert em\varepsilon
\right\rangle +\left\vert me1\right\rangle +\left\vert \varepsilon
1e\right\rangle )].%
\end{array}%
\end{equation}

The formula $\mbox{Tr}_{AB}\left( \left\vert \Psi \right\rangle \left\langle
\Psi \right\vert \right) =\mbox{Tr}_{AC}\left( \left\vert \Psi \right\rangle
\left\langle \Psi \right\vert \right) =\mbox{Tr}_{BC}\left( \left\vert \Psi
\right\rangle \left\langle \Psi \right\vert \right) =I/4$ is still tenable,
and so does braiding any two of $A$, $B$, and $C$ for circles.$\blacksquare $

As stated above, it can be said with certainty that all conclusions
mentioned above are tenable when $A$, $B$, and $C$ are three mutually
orthogonal Latin squares.

\section{The Ising anyon}

\label{Ising}

The algebraic and topological structures in conformal theory lead to the
exotic topological properties of non-Abelian anyons. They are represented in
the framework of topological quantum field theory whose core is a unitary
modular category. The properties of non-Abelian anyons cannot be described
as simply as in the Abelian case. Non-Abelian anyons are those vortices and
gapless fermions whose Chern numbers are odd.

\subsection{Review of the Ising anyon and braiding rules}

\begin{figure}[tbp]
\includegraphics[scale=0.5]{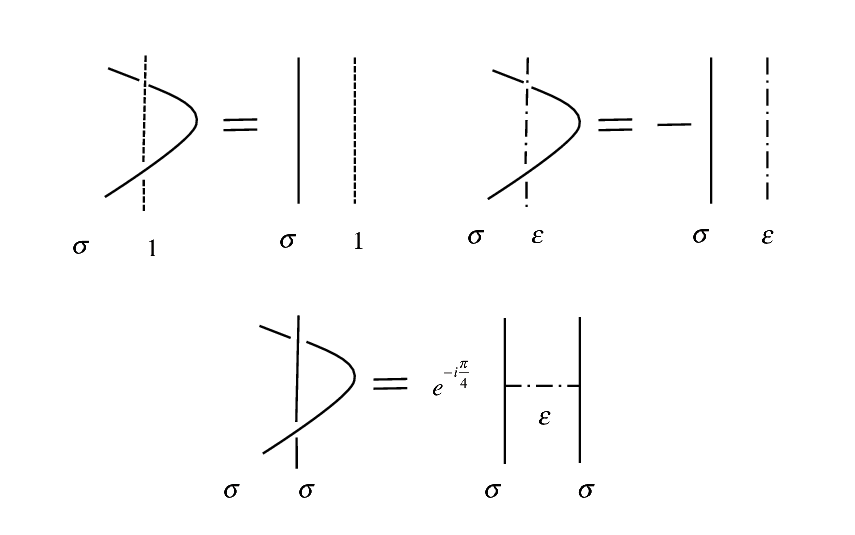}
\caption{The braiding operation of an Ising anyon. A $\protect\sigma $
circling an $\protect\varepsilon $ gives a phase factor $-1$. Two $\protect%
\sigma $'s circling induces $\exp (-i\protect\pi /4)$ and is accompanied
with an exchange of an $\protect\varepsilon $.}
\label{fig3}
\end{figure}

The simplest non-Abelian anyon is called an Ising anyon ($c=1$ is the Chern
number). There are three superselection sectors: the vacuum $1$, the fermion
$\varepsilon $, and the vortex $\sigma $. For non-Abelian anyons, the
topological spin and the Frobenius-Schur indicators of vortices are divided
into eight pieces, $\theta _{\sigma }=\exp (i\pi c/8)$ and $\varkappa
_{\sigma }=(-1)^{(c^{2}-1)/8}$ , and in addition $\theta _{1}=1,:\theta
_{\varepsilon }=-1,:\varkappa _{1}=\varkappa _{\varepsilon }=1$. In Ising
anyon case, $\theta _{\sigma }=\exp (i\pi /8)$ and $\varkappa _{\sigma
}=\varkappa =1$. The fusion rules of non-Abelian anyons are
\begin{equation}
\varepsilon \times \varepsilon =1,\quad \varepsilon \times \sigma =\sigma
,\quad \sigma \times \sigma =1+\varepsilon .
\end{equation}%
According to the definition of braiding (Fig.\ref{fig1}), the braiding rules
of Ising anyons give (Fig.\ref{fig3})
\begin{equation}
\begin{array}{cc}
R_{1}^{\varepsilon \varepsilon }=-1,: & R_{1}^{\sigma \sigma }=\varkappa e^{-%
\frac{i\pi c}{8}}=e^{-\frac{i\pi }{8}}, \\
R_{\sigma }^{\varepsilon \sigma }=R_{\sigma }^{\sigma \varepsilon
}=-i^{c}=-i, & :R_{\varepsilon }^{\sigma \sigma }=\varkappa e^{\frac{i3\pi c%
}{8}}=e^{\frac{i3\pi }{8}}.%
\end{array}
\label{eq:d}
\end{equation}%
Besides an additional phase factor $\exp (-i\pi /4)$, two $\sigma $'s
braiding exchange a fermion $\varepsilon $.

\subsection{Masking protocol of the Ising anyon}

In the Ising anyon case, the space is three dimensional (three
superselection sectors $\left\{ \left\vert 1\right\rangle ,\left\vert
\varepsilon \right\rangle ,\left\vert \sigma \right\rangle \right\} $). The
quantum information can be masked in a $\mathbb{C}^{3}\otimes \mathbb{C}%
^{3}\otimes \mathbb{C}^{3}$ Latin-square construction. We also define the $A$
matrix as
\begin{equation}
A\equiv \left(
\begin{array}{c}
1 \\
1 \\
1%
\end{array}%
\right) \otimes \left(
\begin{array}{ccc}
1 & \varepsilon & \sigma%
\end{array}%
\right) \quad or\quad \left(
\begin{array}{c}
1 \\
\varepsilon \\
\sigma%
\end{array}%
\right) \otimes \left(
\begin{array}{ccc}
1 & 1 & 1%
\end{array}%
\right) .
\end{equation}%
We show the first form of $A$ as an example. The Latin squares in odd
dimensions can be represented by cyclic permutation operators. Here, we
define forward and backward cyclic permutation operators $P_{cf}$ and $%
P_{cb} $ as
\begin{equation}
\begin{array}{ccc}
P_{cf}\left\vert 1,2,3\cdots ,d\right\rangle & = & \left\vert d,1,2\cdots
,d-1\right\rangle , \\
P_{cf}^{2}\left\vert 1,2,3\cdots ,d\right\rangle & = & \left\vert
d-1,d,1\cdots ,d-2\right\rangle , \\
P_{cb}\left\vert 1,2,3\cdots ,d\right\rangle & = & \left\vert 2,3,4\cdots
,d,1\right\rangle , \\
P_{cb}^{2}\left\vert 1,2,3\cdots ,d\right\rangle & = & \left\vert 3,4\cdots
,d,1,2\right\rangle .%
\end{array}
\label{eq:f}
\end{equation}%
So $B$ and $C$ are
\begin{equation}
B=\left(
\begin{array}{c}
1 \\
P_{cf} \\
P_{cf}^{2}%
\end{array}%
\right) \otimes \left(
\begin{array}{ccc}
1 & \varepsilon & \sigma%
\end{array}%
\right) =\left(
\begin{array}{ccc}
1 & \varepsilon & \sigma \\
\sigma & 1 & \varepsilon \\
\varepsilon & \sigma & 1%
\end{array}%
\right) ,
\end{equation}%
\begin{equation}
C=\left(
\begin{array}{c}
1 \\
P_{cb} \\
P_{cb}^{2}%
\end{array}%
\right) \otimes \left(
\begin{array}{ccc}
1 & \varepsilon & \sigma%
\end{array}%
\right) =\left(
\begin{array}{ccc}
1 & \varepsilon & \sigma \\
\varepsilon & \sigma & 1 \\
\sigma & 1 & \varepsilon%
\end{array}%
\right) .
\end{equation}%
Again, through the construction of Eqs. (\ref{eq:b}) and (\ref{eq:c}), an
arbitrary state $\alpha \left\vert 1\right\rangle +\beta \left\vert
\varepsilon \right\rangle +\gamma \left\vert \sigma \right\rangle $ is
mapped into
\begin{equation}
\begin{array}{c}
\left\vert \Psi \right\rangle =\frac{1}{\sqrt{3}}[\alpha (\left\vert
111\right\rangle +\left\vert \varepsilon \varepsilon \varepsilon
\right\rangle +\left\vert \sigma \sigma \sigma \right\rangle ) \\
+\beta (\left\vert 1\sigma \varepsilon \right\rangle +\left\vert \varepsilon
1\sigma \right\rangle +\left\vert \sigma \varepsilon 1\right\rangle ) \\
+\gamma (\left\vert 1\varepsilon \sigma \right\rangle +\left\vert
\varepsilon \sigma 1\right\rangle +\left\vert \sigma 1\varepsilon
\right\rangle )].%
\end{array}
\label{eq:e}
\end{equation}%
In the same way, we can get $\mbox{Tr}_{AB}\left( \left\vert \Psi
\right\rangle \left\langle \Psi \right\vert \right) =\mbox{Tr}_{AC}\left(
\left\vert \Psi \right\rangle \left\langle \Psi \right\vert \right) =%
\mbox{Tr}_{BC}\left( \left\vert \Psi \right\rangle \left\langle \Psi
\right\vert \right) =I/3$, and the masking process is accomplished.

\begin{figure}[tbp]
\includegraphics[scale=0.65]{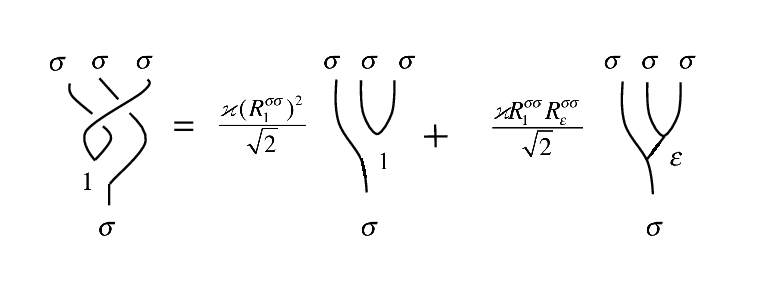}
\caption{The tripartite braiding operation of an Ising anyon.}
\label{fig4}
\end{figure}

The braiding of non-Abelian anyons is a little complicated. Based on Eq. (%
\ref{eq:d}) and Fig.\ref{fig3}, the circling of $\varepsilon $and $\sigma $
gives a phase factor $-1$, and the circling of two $\sigma $'s induces a
phase factor $\exp (-i\pi /4)$ that is accompanied by an exchange of one $%
\varepsilon $. When circling two of the three particles, circling $B$ around
$A$, for example, we have
\begin{equation*}
\begin{array}{c}
\left\vert \Psi \right\rangle =\frac{1}{\sqrt{3}}[\alpha (\left\vert
111\right\rangle -\left\vert \varepsilon \varepsilon \varepsilon
\right\rangle +e^{-i\frac{\pi }{4}}\left\vert \sigma \sigma \sigma
\right\rangle ) \\
+\beta (\left\vert 1\sigma \varepsilon \right\rangle +\left\vert \varepsilon
1\sigma \right\rangle -\left\vert \sigma \varepsilon 1\right\rangle ) \\
+\gamma (\left\vert 1\varepsilon \sigma \right\rangle -\left\vert
\varepsilon \sigma 1\right\rangle +\left\vert \sigma 1\varepsilon
\right\rangle )].%
\end{array}%
\end{equation*}%
In this case, the phases cased by circling cancel out each other in $%
\left\vert \Psi \right\rangle \left\langle \Psi \right\vert $ and $\mbox{Tr}%
_{AB}\left( \left\vert \Psi \right\rangle \left\langle \Psi \right\vert
\right) =\mbox{Tr}_{AC}\left( \left\vert \Psi \right\rangle \left\langle
\Psi \right\vert \right) =\mbox{Tr}_{BC}\left( \left\vert \Psi \right\rangle
\left\langle \Psi \right\vert \right) =I/3$. We can mask the information
too, as does circling of any two particles in other cases. Braiding
(exchange) two adjacent particles, such as $A$ and $B$ or $B$ and $C$, is
similar. For instance, braiding $A$ and $B$ shows
\begin{equation*}
\begin{array}{c}
\left\vert \Psi \right\rangle =\frac{1}{\sqrt{3}}[\alpha (\left\vert
111\right\rangle +R^{\varepsilon \varepsilon }\left\vert \varepsilon
\varepsilon \varepsilon \right\rangle +R^{\sigma \sigma }\left\vert \sigma
\sigma \sigma \right\rangle ) \\
+\beta (\left\vert 1\sigma \varepsilon \right\rangle +\left\vert \varepsilon
1\sigma \right\rangle +R^{\sigma \varepsilon }\left\vert \sigma \varepsilon
1\right\rangle ) \\
+\gamma (\left\vert 1\varepsilon \sigma \right\rangle +R^{\varepsilon \sigma
}\left\vert \varepsilon \sigma 1\right\rangle +\left\vert \sigma
1\varepsilon \right\rangle )].%
\end{array}%
\end{equation*}%
Here, $R^{\sigma \sigma }$ has two forms $R_{1}^{\sigma \sigma }$ or $%
R_{\varepsilon }^{\sigma \sigma }$ according to the fusion rules. Regardless
of the situation, $\mbox{Tr}_{AB}\left( \left\vert \Psi \right\rangle
\left\langle \Psi \right\vert \right) =\mbox{Tr}_{AC}\left( \left\vert \Psi
\right\rangle \left\langle \Psi \right\vert \right) =\mbox{Tr}_{BC}\left(
\left\vert \Psi \right\rangle \left\langle \Psi \right\vert \right) =I/3$.
The masking process can be achieved. The exchange of $A$ and $C$ proceeds in
two stages: One is a tripartite braiding (Fig.\ref{fig4}), and the other is
the exchange of $B$ and $C$.
The tripartite braiding is divided into two parts with different
probabilities (Fig. \ref{fig4}). So the mapping goes to
\begin{equation}
\begin{array}{ccc}
\left\vert \Psi \right\rangle & \text{=} & \frac{1}{\sqrt{3}}[\alpha
(\left\vert 111\right\rangle +(R^{\varepsilon \varepsilon })^{3}\left\vert
\varepsilon \varepsilon \varepsilon \right\rangle \\
& \text{+} & \frac{\varkappa (R_{1}^{\sigma \sigma })^{2}}{\sqrt{2}}%
\left\vert \sigma \sigma \sigma \right\rangle _{1}+\frac{\varkappa
R_{1}^{\sigma \sigma }R_{\varepsilon }^{\sigma \sigma }}{\sqrt{2}}\left\vert
\sigma \sigma \sigma \right\rangle _{\varepsilon }) \\
& \text{+} & \beta (R^{\sigma \varepsilon }\left\vert 1\sigma \varepsilon
\right\rangle +R^{\varepsilon \sigma }\left\vert \varepsilon 1\sigma
\right\rangle +R^{\sigma \varepsilon }\left\vert \sigma \varepsilon
1\right\rangle ) \\
& \text{+} & \gamma (R^{\varepsilon \sigma }\left\vert 1\varepsilon \sigma
\right\rangle +R^{\varepsilon \sigma }\left\vert \varepsilon \sigma
1\right\rangle +R^{\sigma \varepsilon }\left\vert \sigma 1\varepsilon
\right\rangle )].%
\end{array}%
\end{equation}%
It is worth noting that the states $\left\vert \sigma \sigma \sigma
\right\rangle _{1}$ and $\left\vert \sigma \sigma \sigma \right\rangle
_{\varepsilon }$ are mutually orthogonal. Then, as mentioned above, the
masking process is invariant under the exchange of two adjacent particles.
Substituting the values of those braiding operations, we also conclude the
partial traces $\mbox{Tr}_{AB}\left( \left\vert \Psi \right\rangle
\left\langle \Psi \right\vert \right) =\mbox{Tr}_{AC}\left( \left\vert \Psi
\right\rangle \left\langle \Psi \right\vert \right) =\mbox{Tr}_{BC}\left(
\left\vert \Psi \right\rangle \left\langle \Psi \right\vert \right) =I/3$.
This also confirms that the braiding and circling of Ising anyons are
actually on extended hyperdisks in the anyonic space of the Kitaev model,
and thus gives an extended support of Ref. \cite{z0}. Similar to the above
Abelian part, for Ising anyons, the ideal situation is that matrices $A$, $B$%
, and $C$ are three mutually orthogonal Latin squares. In this ideal
situation, all of the above discussed are still established.

\subsection{Masking based on teleportation}

It has been proved that there exist certain correlations between quantum
information masking and teleportation. In Ref. \cite{z}, Shang \emph{et al.}
point out that teleportation is a process which masks the information first
during the transference. Reference \cite{b7} gives a quantum teleportation
scheme using Ising anyons. Here, we demonstrate quantum information masking
in a teleportation way in the Kitaev Ising anyon model. Suppose the state to
be teleported is $\left\vert \chi \right\rangle =\alpha \left\vert
1\right\rangle +\beta \left\vert \varepsilon \right\rangle +\gamma
\left\vert \sigma \right\rangle $ ($\left\vert \alpha \right\vert
^{2}+\left\vert \beta \right\vert ^{2}+\left\vert \gamma \right\vert ^{2}=1$%
). Alice and Bob share an entangled quantum channel
\begin{equation}
\left\vert \varphi \right\rangle _{23}=\frac{1}{\sqrt{3}}(\left\vert
11\right\rangle +\left\vert \varepsilon \varepsilon \right\rangle
+\left\vert \sigma \sigma \right\rangle ).
\end{equation}%
Alice has the first and the second particles, and Bob holds the third one.
The state of the whole system is
\begin{equation}
\left\vert \Psi \right\rangle _{123}=\frac{1}{\sqrt{3}}(\alpha \left\vert
1\right\rangle +\beta \left\vert \varepsilon \right\rangle +\gamma
\left\vert \sigma \right\rangle )_{1}(\left\vert 11\right\rangle +\left\vert
\varepsilon \varepsilon \right\rangle +\left\vert \sigma \sigma
\right\rangle )_{23}.
\end{equation}%
With the help of Eq. (\ref{eq:f}), the above equation is mapped to
\begin{equation}
\begin{array}{ccc}
\left\vert \Psi ^{^{\prime }}\right\rangle _{123} & = & \frac{1}{\sqrt{3}}%
(\alpha \left\vert 1\right\rangle +\beta \left\vert \varepsilon
\right\rangle +\gamma \left\vert \sigma \right\rangle )_{1}(\left\vert
11\right\rangle \\
& + & P_{cf2}P_{cb3}\left\vert \varepsilon \varepsilon \right\rangle
+P_{cf2}^{2}P_{cb3}^{2}\left\vert \sigma \sigma \right\rangle )_{23},%
\end{array}%
\end{equation}%
where $P_{cf2}$ means the forward cyclic permutation operator which acts on
the second particle. The matrices of the second and the third particles are
mutually orthogonal Latin squares. The quantum information can be masked in
the partial systems. After the operation,
\begin{equation}
\begin{array}{ccc}
\left\vert \Psi ^{^{\prime }}\right\rangle _{123} & = & \frac{1}{3}%
[(\left\vert \chi _{1}\right\rangle )_{12}(\alpha \left\vert 1\right\rangle
+\beta \left\vert \varepsilon \right\rangle +\gamma \left\vert \sigma
\right\rangle )_{3} \\
& + & (\left\vert \chi _{2}\right\rangle )_{12}(\alpha \left\vert
1\right\rangle +\beta \omega ^{2}\left\vert \varepsilon \right\rangle
+\gamma \omega \left\vert \sigma \right\rangle )_{3} \\
& + & (\left\vert \chi _{3}\right\rangle )_{12}(\alpha \left\vert
1\right\rangle +\beta \omega \left\vert \varepsilon \right\rangle +\gamma
\omega ^{2}\left\vert \sigma \right\rangle )_{3}],%
\end{array}%
\end{equation}%
\begin{equation*}
\begin{array}{ccc}
(\left\vert \chi _{1}\right\rangle )_{12} & = & \left\vert 11\right\rangle
+\left\vert 1\varepsilon \right\rangle +\left\vert 1\sigma \right\rangle
+\left\vert \varepsilon 1\right\rangle +\left\vert \varepsilon \varepsilon
\right\rangle \\
& + & \left\vert \varepsilon \sigma \right\rangle +\left\vert \sigma
1\right\rangle +\left\vert \sigma \varepsilon \right\rangle +\left\vert
\sigma \sigma \right\rangle ,%
\end{array}%
\end{equation*}%
\begin{equation*}
\begin{array}{ccc}
(\left\vert \chi _{2}\right\rangle )_{12} & = & \left\vert 11\right\rangle
+\omega \left\vert 1\varepsilon \right\rangle +\omega ^{2}\left\vert 1\sigma
\right\rangle +\omega ^{2}\left\vert \varepsilon 1\right\rangle +\left\vert
\varepsilon \varepsilon \right\rangle \\
& + & \omega \left\vert \varepsilon \sigma \right\rangle +\omega \left\vert
\sigma 1\right\rangle +\omega ^{2}\left\vert \sigma \varepsilon
\right\rangle +\left\vert \sigma \sigma \right\rangle ,%
\end{array}%
\end{equation*}%
\begin{equation*}
\begin{array}{ccc}
(\left\vert \chi _{3}\right\rangle )_{12} & = & \left\vert 11\right\rangle
+\omega ^{2}\left\vert 1\varepsilon \right\rangle +\omega \left\vert 1\sigma
\right\rangle +\omega \left\vert \varepsilon 1\right\rangle +\left\vert
\varepsilon \varepsilon \right\rangle \\
& + & \omega ^{2}\left\vert \varepsilon \sigma \right\rangle +\omega
^{2}\left\vert \sigma 1\right\rangle +\omega \left\vert \sigma \varepsilon
\right\rangle +\left\vert \sigma \sigma \right\rangle ,%
\end{array}%
\end{equation*}

where $\omega =e^{\frac{i2\pi }{3}}$. The information is transferred from
Alice to Bob. This protocol demonstrates the teleportation process using a
masking mapping.

\section{summary}

In the task of quantum masking, the information information encoded in a
single system is distributed to the correlations among a composite system.
Modi \emph{et al.} \cite{1} pointed out that it is an impossible task in a
bipartite system, while it can be achieved when more participants are
allowed in the masking process.
This phenomenon originated from the linearity, unitarity, and entanglement
in quantum mechanics. 

Here, we adopt exotic anyons in two-dimensional condensed-matter systems as
the platform to realize multipartite masking scenarios. The anyonic space is
quite different from the normal Hilbert space, and the framework of anyonic
algebra is in a unitary braided fusion category. So how to mask information
in the anyonic space and the maskable construction is worth researching and
is conceivably complicated. 
Based on Theorem $2$ in Ref. \cite{2}, we present the Latin-square masking
protocols both in Abelian and non-Abelian anyons.
We may safely draw the conclusion that the maskable constructions in anyonic
space are extended hyperdisks, and the anyonic quantum entanglement which
originates from the braiding operations is the source of the quantum
information masking. We also realize quantum information masking in a
teleportation way in the Kitaev Ising model, to reveal the relation between
masking and teleportation in non-Abelian anyons.

More protocols of other Abelian and non-Abelian anyons cases and more
realizations on the unitary evolution in quantum field systems will be
discussed in our future works. Storing the information in the correlation is
a material difference between a quantum computer and a classical computer.
It goes without saying that quantum information masking is of great
significance to the development of quantum computers and its related
evidence collection in the future.

\section{Acknowledgment}

The research was supported by the Fundamental Research Funds for the Central
Universities, China No. 2022JKF02024, the National Natural Science
Foundation of China No. 11675119, and the National Natural Science
Foundation of China No. 62106033.

\end{document}